\documentclass[aps,prl,twocolumn,showpacs,superscriptaddress,groupedaddress]{revtex4-2}  
\usepackage{graphicx}  
\usepackage{dcolumn}   
\usepackage{bm}        
\usepackage{amssymb}   
\usepackage[errorshow]{tracefnt}
\usepackage{csquotes}
\usepackage{longtable}
\usepackage{tabularx}
\usepackage{multirow}
\usepackage{tikz}
\usetikzlibrary{arrows,snakes,shapes}
\usetikzlibrary{plotmarks}
\usepackage{pgfplots}
\usepackage{booktabs}
\usepackage{threeparttable} 
\usepackage{rotating}
\usepackage{xcolor}
\usepackage{epigraph} 
\usepackage{amsmath}
\usepackage{scalefnt}

\begin{document}

\title{The $B(E2)$ anomaly: Evidence for a low-lying mixed-symmetry collective excitation mode}
\author{Bo~Cederwall}
\affiliation{KTH Royal Institute of Technology, 10691 Stockholm, Sweden}
\thanks{Corresponding author}
\email{bc@kth.se}
\author{Chong~Qi}
    \affiliation{KTH Royal Institute of Technology, 10691 Stockholm, Sweden}
\vskip 0.25cm

\date{\today}

\begin{abstract}{
Exceptionally low values of the ratio of electric quadrupole transition rates, $B_{4/2}\equiv 
B(E2;4^+_1\rightarrow2^+_1)/B(E2;2^+_1\rightarrow0^+_{\mathrm{gs}})<1$, have been observed in neutron-deficient nuclei near $N\approx94$ (W, Os, Pt) and $N\approx62$ (Te, Xe) with few and comparable numbers of valence nucleons outside closed shells. Remarkably, the suppressed $B_{4/2}$ ratios coincide with low-lying energy level patterns characteristic of collective motion. Standard approaches, including large-scale shell model, collective models, and density functional theory, fail to reproduce this behavior, commonly referred to as the $B_{4/2}$ (or $B(E2)$) anomaly.
Recent work has reproduced the effect in selected Pt and Os isotopes via mapping a triaxial rotor Hamiltonian onto the interacting boson model (IBM), attributing it to triaxial rotational motion. However, this interpretation is unexpected as collectivity typically emerges first through vibrational modes with increasing valence nucleon number along isotopic chains.
Here, we address this discrepancy using an extended IBM Hamiltonian across nuclei exhibiting the anomaly, benchmarked against large-scale shell model calculations, and propose that the $B(E2)$ anomaly arises from a low-lying mixed-symmetry collective mode that bridges single-particle and collective dynamics.
}
\end{abstract}

\maketitle

\section{Introduction}
Quantum many-body systems often exhibit collective degrees of freedom, i.e., coherent motion emerging from strongly interacting constituents~\cite{Otsuka2019}. Atomic nuclei provide a paradigmatic example, where interacting neutron and proton Fermi systems give rise to quantum phase transitions, shape coexistence, and collective excitations with characteristic regularity.
Dynamical symmetries provide a unifying framework for this behavior by defining analytic limits of the nuclear many-body problem and organizing the emergence of collectivity. In algebraic approaches such as the interacting boson model (IBM)~\cite{ibm,casten}, these symmetries correspond to distinct structural regimes [U(5), O(6), SU(3)] and associated excitation modes. Deviations from symmetry limits, including the emergence of mixed-symmetry states, probe the interplay between single-particle and collective degrees of freedom and are therefore particularly sensitive to changes in nuclear structure.
Empirically, nuclei several nucleons away from magic numbers exhibit collective behavior characterized in even–-even systems by excitation energy ratios $R_{4/2}=E_{4^+_1}/E_{2^+_1}\gtrsim 2$ and electric quadrupole transition strengths that increase with angular momentum~\cite{bohrmottelson}. In this regime, the ratio
$B_{4/2}\equiv \frac{B(E2; 4^+_1\rightarrow 2^+_1)}{B(E2; 2^+_1\rightarrow 0^+_{\mathrm{gs}})}$
exceeds unity, with characteristic limits $B_{4/2}=10/7$ for rotational motion (Alaga rule) and $B_{4/2}=2$ for harmonic vibrations~\cite{casten,heyde}. These features reflect coherent multi-particle configurations and evolve smoothly along isotopic chains, from weakly collective spherical nuclei near closed shells to well-deformed systems with rotational spectra near mid-shell.
However, systematic observations of nuclei exhibiting $B_{4/2}<1$ challenge this picture. Such anomalous ratios have been identified in neutron-deficient regions around $N\approx94$ (W, Os, Pt) and $N\approx62$ (Te, Xe), where they coincide with low-lying level structures consistent with collective motion, including $R_{4/2}>2$. This behavior lies outside the expectations of standard collective models, large-scale shell model calculations, and density functional approaches, and is commonly referred to as the $B(E2)$ anomaly.

Recent work has suggested a possible interpretation in terms of triaxial rotational dynamics within an IBM framework~\cite{zhang2022,zhang2022,zhang2024,teng2025,pan2024}. This is nontrivial, as collectivity is generally expected to develop first through vibrational modes with increasing valence nucleon number. In this work, we revisit the origin of the $B(E2)$ anomaly using an extended IBM Hamiltonian, bench marked against large-scale shell model calculations, suggesting that the anomaly arises from the emergence of a low-lying mixed-symmetry collective mode, providing a microscopic link between single-particle and collective dynamics.

\section{Discussion}
With the improvements of experimental instrumentation and techniques, electromagnetic transition strengths are becoming increasingly accessible for a large number of nuclei far from stability. 
The importance of such measurements for understanding the underlying mechanisms behind the emergence of collective nuclear excitations is highlighted by an increasing number of exceptions from the expected $B_{4/2}$ values that have been observed in even-even nuclei, such as $^{112}$Te~\cite{doncel2017}, $^{114}$Te~\cite{cakirli2004,moller2005}, $^{112}$Xe~\cite{juradophd}, $^{114}$Xe~\cite{deangelis2002}, $^{166}$W~\cite{saygi2017}, $^{168}$Os~\cite{grahn2016}, $^{170}$Os~\cite{goasduff2019}  and $^{172}$Pt~\cite{cederwall2018}, despite the collective character of their energy level structures and the distance from closed shells.
If we generalize the definition of the $B_{4/2}$ ratio to include nuclei with odd neutron numbers ($N$) and/or odd proton numbers ($Z$), as: $B_{4/2} = B(E2;J+4 \to J+2)/ B(E2;J+2 \to J)$, where $J$, is the angular momentum of the single-particle configuration forming the ``band head'' of the excited structure of interest, the same ``anomaly'' is then observed in $^{169}$Os ~\cite{zhang2021}, $^{167}$Os, and $^{165}$W with measured ratios $\frac{B(E2;21/2^+ \to 17/2^+)}{B(E2;17/2^+ \to 13/2^+)} = $ 0.79(16)~\cite{zhang2021}, 0.49(10)~\cite{zanon2024} and 0.59(7)~\cite{Milanovic2026}, respectively.
Similarly to the now well-established region of anomalous cases located in the neutron deficient isotopes of the W-Os-Pt transition metals, the Te-Xe nuclei are well separated from the closest magic neutron number at $N=50$ and are therefore, \textit{a priori}, expected to develop collective excitations and $B_{4/2} > 1$ with increasing neutron number. 
 The semi-magic $^{108,112,114}$Sn~\cite{siciliano2020,kundu2021,jonsson1981,gableske2001} also exhibit $B_{4/2} <1$ but for these cases there is reason to allow for that partial seniority symmetry within the $\nu h_{11/2}$ intruder subshell or mixed $\nu (d_{5/2},g_{7/2})$ natural-parity subshells may play a role in reducing the $B_{4/2}$ ratios.
 Furthermore, several of the nuclei in the lighter mass region are also accessible for LSSM 
configuration interaction calculations, which we take advantage of in the present work.

\begin{figure}[htbp]
\begin{tikzpicture}[scale=0.9]
\tikzstyle{every node}=[font=\small]
\def\ep{1.5}
\def\en{0.8}
\begin{axis}[
xtick={6,8,10,12,14,16,18,20,22,24},
ytick={0,0.2,0.4,0.6,0.8,1.0,1.2,1.4,1.6,1.8,2.0,2.2,2.4,2.6,2.8,3.0,3.2,3.4},
xticklabels={},
xmin=4,xmax=25,ymin=1,ymax=3.5,ytickmax=3.4, minor y tick num = 1,
ylabel=$R_{4/2}$,
height=6cm,width=0.99\columnwidth,name=axisE42,
legend style={at={(1.005,0.30)}}
]
\addplot[magenta, mark=diamond*] plot[
only marks,
mark size=4pt,
error bars/.cd,
y dir=both, y explicit,
x dir=both, x explicit
]
coordinates {
(12.2,2.33) +- (0, 0.0)
(16.0,2.14) +- (0, 0.0)
(20.0,2.68) +- (0, 0.0)
(22.0,2.71) +- (0, 0.0)
};
\addlegendentry{Pt, experiment}
\addplot+[magenta, mark=triangle] plot[
only marks,
mark size=5pt
]
coordinates {
(11.8,2.336) 
};
\addlegendentry{Pt, IBM-1 + H$_{MS}$} 
\addplot[black, mark=square*] plot[
only marks,
mark size=3pt,
error bars/.cd,
y dir=both, y explicit,
x dir=both, x explicit
]
coordinates {
(9,2.50) +- (0,0.0)
(10.2,2.51) +- (0, 0.0)
(11,2.71) +- (0,0.0)
(12.2,2.61) +- (0, 0.0)
(13,2.77) +- (0,0.0)
(14,2.66) +- (0, 0.0) 
(15,2.67) +- (0,0.0) 
(16,2.66) +- (0, 0.0)
(22,3.1) +- (0, 0.0)
(24,3.15) +- (0, 0.0)
};
\addlegendentry{Os, experiment}
\addplot+[black, mark=triangle] plot[
only marks,
mark size=5pt
]
coordinates {
(9.8,2.513) 
(11,2.81) 
(11.8,2.62) 
(13.1,2.95) 
};
\addlegendentry{Os, IBM1 + H$_{MS}$} 
\addplot[green, mark=o] plot[
only marks,
mark size=4pt,
error bars/.cd,
y dir=both, y explicit,
x dir=both, x explicit
]
coordinates {
(9.9,2.68) +- (0, 0.0)
(11.9,2.82) +- (0, 0.0)
(13.9,2.95) +- (0, 0.0)
(21.9,3.1) +- (0, 0.0)
};
\addlegendentry{W, experiment}
\addplot {x*x};
\draw [thick, dashed, draw=black] 
(axis cs: 4,3.33) -- (axis cs: 25,3.33)
        node[pos=0.4, below] {$symm.~rotor$};
\addplot {x*x};
\draw [thick, dashed, draw=black] 
        (axis cs: 4,2.5) -- (axis cs: 25,2.5)
        node[pos=0.6, below] {$asym.~rotor$};
\draw [thick, dashed, draw=black] 
        (axis cs: 4,2.0) -- (axis cs: 25,2.0)
        node[pos=0.4, below] {$vibr.~limit$};
\end{axis}
\end{tikzpicture}
\begin{tikzpicture}[scale=0.9]
\def\ep{1.5}
\def\en{0.8}
\begin{axis}[
xtick={6,8,10,12,14,16,18,20,22,24},
xticklabels={3,4,5,6,7,8,9,10,11,12},
ytick={0,0.2,0.4,0.6,0.8,1.0,1.2,1.4,1.6,1.8,2.0,2.2,2.4},
xmin=4,xmax=25,ymin=0,ymax=2.5,ytickmax=2.4, minor y tick num = 1,
xlabel=$N_\nu$,
ylabel=$B_{4/2}$,
height=10cm,width=0.99\columnwidth,name=axisXeBE2,
]

\addplot[magenta, mark=diamond*] plot[
only marks,
mark size=4pt,
error bars/.cd,
y dir=both, y explicit,
x dir=both, x explicit
]
coordinates {
(12.0,0.557401455) +- (0, 0.279692766)
(16.2,1.89) +- (0, 0.34)
(20.0,1.42) +- (0, 0.10)
(22.0,1.69) +- (0, 0.115)
};
\addplot+[magenta, mark=triangle] plot[
only marks,
mark size=5.5pt
]
coordinates {
(11.9,0.552) 
};
\addplot[black, mark=square*] plot[
only marks,
mark size=3pt,
error bars/.cd,
y dir=both, y explicit,
x dir=both, x explicit
]
coordinates {
(10.2,0.32298865) +- (0, 0.216638728)
(11,0.79) +- (0,0.16)
(12.2,0.39) +- (0, 0.12)             
(13,0.96) +- (0,0.15)
(14,1.5169146) +- (0, 0.239550327) 
(15,1.78) +- (0,0.38)
(15.8,1.89619626) +- (0, 0.304299417)
(22.2,1.4) +- (0, 0.2)
(24,1.39) +- (0, 0.07)
};
\addplot+[black, mark=triangle] plot[
only marks,
mark size=5.5pt
]
coordinates {
(9.8,0.351) 
(11.1,0.69) 
(11.8,0.41) 
(13.1,0.76) 
};
\addplot[green, mark=o] plot[
only marks,
mark size=4pt,
error bars/.cd,
y dir=both, y explicit,
x dir=both, x explicit
]
coordinates {
(9.0,0.59) +- (0, 0.07)
(9.8,0.337018871) +- (0, 0.071658498)
(11.9,1.120060774) +- (0, 0.318269808)
(13.9,1.428753993) +- (0, 0.164209786)
(21.8,1.409154291) +- (0, 0.216272963)
};
\addplot {x*x};
\draw [thick, dotted, draw=black] 
        (axis cs: 4,1.43) -- (axis cs: 25,1.43)
        node[pos=0.2, below] {$Alaga$};
\addplot {x*x};
\draw [thick, dashed, draw=black] 
        (axis cs: 4,2.0) -- (axis cs: 25,2.0)
        node[pos=0.2, above] {$vibr. limit$};
\end{axis}
\end{tikzpicture}
\caption{(Color online) Experimental excitation energy ratios, $R_{4/2}$, and ratios of reduced E2 transitions rates, $B_{4/2}$, (when available) for tungsten, osmium, and platinum isotopes with neutron numbers $88 \leq N \leq 106$ (bottom), as a function of the number of valence neutron pairs, $N_\nu$. The experimental data are taken from Refs.~\cite{zhang2021,zanon2024,milanovic2025,grahn2016,saygi2017,cederwall2018,goasduff2019}, and \cite{nudat}. The theoretical IBM predictions are from Refs.~\cite{zhang2022,zhang2024,teng2025,pan2024}. Some data points have been somewhat displaced horizontally for clarity.}
\label{fig:E42_B42_exp_1}
\end{figure}
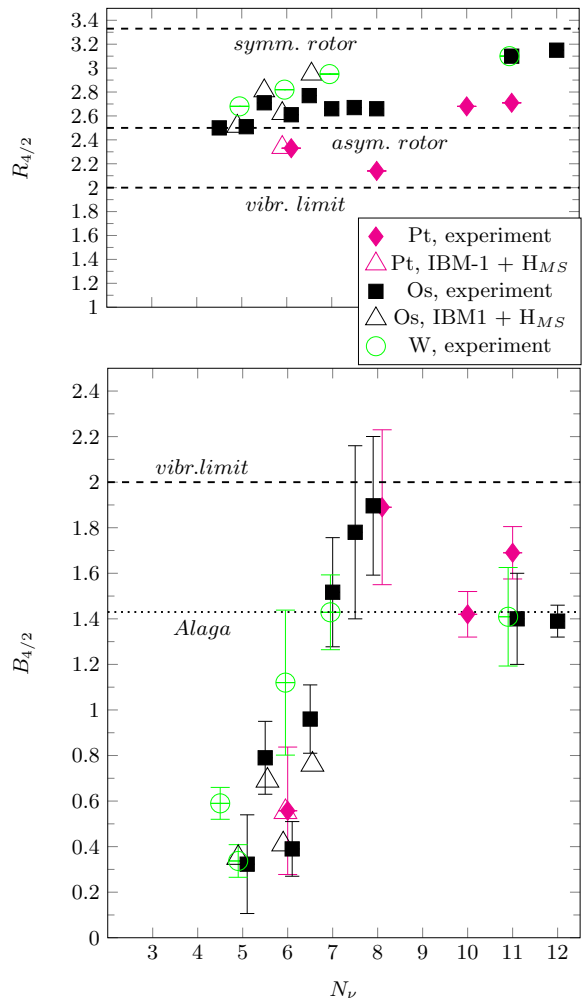

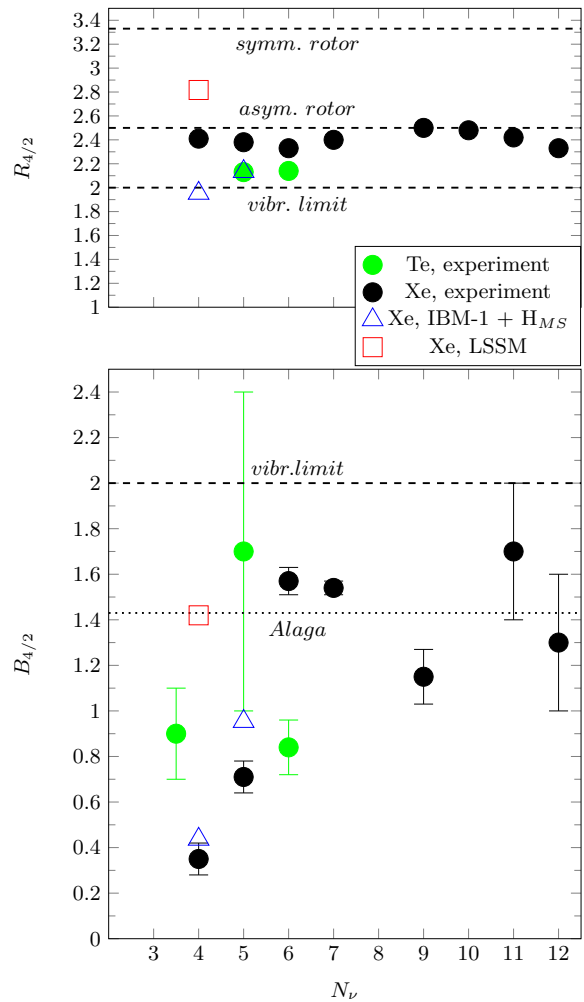
\begin{figure}[htbp]
\begin{tikzpicture}[scale=0.9]
\def\ep{1.5}
\def\en{0.8}
\begin{axis}[
xtick={6,8,10,12,14,16,18,20,22,24},
ytick={0,0.2,0.4,0.6,0.8,1.0,1.2,1.4,1.6,1.8,2.0,2.2,2.4,2.6,2.8,3.0,3.2,3.4},
xticklabels={},
xmin=4,xmax=25,ymin=1,ymax=3.5,ytickmax=3.4, minor y tick num = 1,
ylabel=$R_{4/2}$,
height=6cm,width=0.99\columnwidth,name=axisE42,
legend style={at={(1.00,0.205)}}
]
%
\addplot[green, mark=*] plot[
only marks,
mark size=4pt,
error bars/.cd,
y dir=both, y explicit,
x dir=both, x explicit
]
coordinates {
(7,0.9) +- (0,0.0) 
(10,2.13) +- (0, 0.0)
(12,2.14) +- (0, 0.0)
};
\addlegendentry{Te, experiment}
\addplot[black, mark=*] plot[
only marks,
mark size=4pt,
error bars/.cd,
y dir=both, y explicit,
x dir=both, x explicit
]
coordinates {
(8,2.41) +- (0, 0.0)
(10,2.38) +- (0, 0.0)
(12,2.33) +- (0, 0.0)
(14,2.40) +- (0, 0.0)
(18,2.50) +- (0,0.0) 
(20,2.48) +- (0,0.0)
(22,2.42) +- (0,0.0)
(24,2.33) +- (0,0.0)
};

\addlegendentry{Xe, experiment }

\addplot[blue, mark=triangle] plot[
only marks,
mark size=5pt
]
coordinates {
(8,1.953) 
(10,2.135) 
};
\addlegendentry{Xe, IBM-1 + H$_{MS}$} 
\addplot[color=red, mark=square] plot[
only marks,  
mark size=4pt
]
coordinates {
(8,2.817) 
};
\addlegendentry{Xe, LSSM}
%
%

%
\addplot {x*x};
\draw [thick, dashed, draw=black] 
(axis cs: 4,3.33) -- (axis cs: 25,3.33)
        node[pos=0.4, below] {$symm.~rotor$};
\addplot {x*x};
\draw [thick, dashed, draw=black] 
        (axis cs: 4,2.5) -- (axis cs: 25,2.5)
        node[pos=0.4, above] {$asym.~rotor$};
\draw [thick, dashed, draw=black] 
        (axis cs: 4,2.0) -- (axis cs: 25,2.0)
        node[pos=0.4, below] {$vibr.~limit$};
\end{axis}
\end{tikzpicture}

\begin{tikzpicture}[scale=0.9]
\def\ep{1.5}
\def\en{0.8}
\begin{axis}[
xtick={6,8,10,12,14,16,18,20,22,24},
xticklabels={3,4,5,6,7,8,9,10,11,12},
ytick={0,0.2,0.4,0.6,0.8,1.0,1.2,1.4,1.6,1.8,2.0,2.2,2.4},
xmin=4,xmax=25,ymin=0,ymax=2.5,ytickmax=2.4, minor y tick num = 1,
xlabel=$N_\nu$,
ylabel=$B_{4/2}$,
height=10cm,width=0.99\columnwidth,name=axisXeBE2,
]
%

\addplot[green, mark=*] plot[
only marks,
mark size=4pt,
error bars/.cd,
y dir=both, y explicit,
x dir=both, x explicit
]
coordinates {
(7,0.9) +- (0,0.2) 
(10,1.7) +- (0, 0.7)
(12,0.84) +- (0, 0.12)
};
\addplot[black, mark=*] plot[
only marks,
mark size=4pt,
error bars/.cd,
y dir=both, y explicit,
x dir=both, x explicit
]
coordinates {
(8,0.35) +- (0, 0.07)
(10,0.71) +- (0, 0.07)
(12,1.57) +- (0, 0.06)
(14,1.54) +- (0, 0.03)
(18,1.15) +- (0,0.12) 
(20,3.8) +- (0,0.4)
(22,1.7) +- (0,0.3)
(24,1.3) +- (0,0.3)
};
%
%
%
%
%
\addplot+[blue, mark=triangle,
  error bars/.cd,
    y dir=both,
    y explicit
] plot[
  only marks,
  mark size=5pt
]
coordinates {
  (8, 0.45)  +- (0, 0.17)  
  (10, 0.70) +- (0, 0.08)  
  (12, 1.55) +- (0, 0.05)  
};
%
\addplot+[red, mark=square] plot[
only marks, 
mark size=4pt
]
coordinates {
(8,1.4197) 
};

\addplot {x*x};
\draw [thick, dotted, draw=black] 
        (axis cs: 4,1.43) -- (axis cs: 25,1.43)
        node[pos=0.4, below] {$Alaga$};
\addplot {x*x};
\draw [thick, dashed, draw=black] 
        (axis cs: 4,2.0) -- (axis cs: 25,2.0)
        node[pos=0.4, above] {$vibr. limit$};
\end{axis}
\end{tikzpicture}
\caption{(Color online)
Experimental and calculated excitation energy ratios  ($R_{4/2}=E_{4^{+}_1}/E_{2^{+}_1}$ and B(E2:$4^+_1\rightarrow 2^+_{1}$)/B(E2:$2^+_1\rightarrow 0^+_{gs}$) ratios for tellurium and xenon isotopes with neutron number $57 \leq N \leq 64$ as a function of the number of valence neutron pairs, $N_\nu$. The experimental data are taken from Refs.~\cite{deangelis2002,juradophd,siciliano2020,doncel2017}, and from ~\cite{nudat}. The theoretical IBM and LSSM calculations are from this work, see text for details.
}
\label{fig:E42_B42_exp_2}
\end{figure}

Figure~\ref{fig:E42_B42_exp_1} shows experimental excitation energy ratios, $R_{4/2} \equiv E_{4^{+}_1}/E_{2^{+}_1}$
for even-mass isotopes or $R_{4/2} \equiv (E_{21/2^{+}}-E_{13/2^{+}})/(E_{17/2^{+}}-E_{13/2^{+}})$ for the $\nu i_{13/2^{+}}$ bands in odd-mass nuclei 
(top panel) and $B_{4/2} \equiv \frac{B(E2; 4^+_1\rightarrow 2^+_1)}{B(E2; 2^+_1\rightarrow 0^+_{\mathrm{gs}})}$ for even--even isotopes or $B_{4/2} \equiv \frac{B(E2; 21/2^+_1\rightarrow 17/2^+_1)}{B(E2; 17/2^+_1\rightarrow 13/2^+_1)}$ for $\nu i_{13/2^{+}}$ bands in odd-mass nuclei (bottom panel) for W-Os-Pt isotopes.
Figure~\ref{fig:E42_B42_exp_2}
shows similar quantities for several Te--Xe isotopes.
The data are shown together with dashed and dotted lines indicating collective-model limits for comparison. In particular the W-Os-Pt region, where the experimental data are more abundant, a clear trend is observed for the $B_{4/2}$ ratios as a function of the number of neutron pairs, $N_\nu$. Starting from the lowest numbers of valence neutron pairs, $N_\nu \approx 5$, with $B_{4/2} \lesssim 0.5$, the $B_{4/2}$ ratios increase steadily towards the harmonic vibrational limit and thereafter approach the symmetric rotor limit for $N_\nu \gtrsim 9$. Hence, after the range of $N_\nu$ with "anomalous" $B_{4/2}$ ratios, a normal evolution of collective excitations; vibrations followed by increasingly collective rotational excitations, follow.

\subsection{Models}
Established theoretical models have so far, except in a few isolated cases as discussed below, been unsuccessful in elucidating the mechanisms behind anomalous $B_{4/2}$ ratios.  After the failure of collective, ``geometrical" models and mean field models to reproduce the effect, it was natural to apply the group-theoretical or ``algebraic" approach of the IBM. 
For the case of $^{168}$Os, Grahn \emph{et al.}~\cite{grahn2016} carried out calculations of excited-state energies and absolute $B(E2)$ values within the proton-neutron interacting boson model (IBM-2)~\cite{otsuka1978} for which the parameters of IBM-2 were constrained by SkM* energy density functional calculations and concluded, however, that the experimental data could not be accounted for in this way. Recently, state-of-the-art IBM-2 with configuration mixing (IBM-2 CM) calculations based on energy density functional (EDF) and self-consistent mean-field (SCMF) calculations were carried out by Nomura~\cite{Nomura2026}. The configuration mixing in the case of the IBM-2 CM calculations involved mixing of the normal and intruder states. Similarly to the case of $^{168}$Os~\cite{grahn2016}, neither the standard IBM-2 nor the IBM-2 CM calculations were able to reproduce the observed "anomalous" $B_{4/2}$ ratio in the tellurium isotope $^{114}$Te~\cite{Nomura2026}. These calculations show that also when neutron-proton interactions and configuration mixing are taken into account the collective electromagnetic properties stay regular in the case of a standard IBM Hamiltonian.

In the conventional description of collective states within the IBM, three different dynamical symmetry limits are realized; U(5), O(6), and SU(3)~\cite{ibm,warner1983,jolie2001}. 
They originate from the following three chains of the U(6) group~\cite{ibm} :
\begin{equation}
    U(6) \supset U(5) \supset O(5) \supset O(3),
\end{equation}
\begin{equation}
    U(6) \supset SU(3) \supset O(3),
\end{equation}
\begin{equation}
    U(6) \supset O(6) \supset O(5) \supset O(3).
\end{equation}
The dynamical symmetry associated with each group chain corresponds
to a characteristic collective mode,  i.e. axially-symmetric deformed rotor (SU(3)), $\gamma$-unstable rotor (O(6)), and surface vibrations around a spherical shape (U(5)).

A common variety of the standard IBM Hamiltonian uses the ``consistent-Q" formalism in which it can be written~\cite{warner1983}: 
\begin{equation}
    H_{CQ} = \epsilon \, \hat{n}_d + \kappa \frac{1}{N} \hat{Q}_\chi \cdot \hat{Q}_\chi,
\end{equation}
with 
\begin{equation}
    \hat{n}_d = d^\dagger \cdot \tilde{d}, \quad \text{and} \quad \hat{Q}_\chi^u = (d^\dagger s + s^\dagger \tilde{d})^{(2)}_u + \chi (d^\dagger \times \tilde{d})^{(2)}_u,
\end{equation}
where $\epsilon$, $\kappa$, and $\chi$ are real parameters, and $N$ is the total boson number.
In this formalism, the E2 transition operator is defined as 
\(
T(E2) = e_B \hat{Q}_\chi
\)
, where \(\hat{Q}_\chi\)
is the quadrupole operator in the Hamiltonian and 
\(e_B\) denotes the effective boson charge.

We note that the three dynamical-symmetry limits of $H_{CQ}$ are realized only at the special points $\chi=0$ (O(6)), $\chi=\pm\sqrt{7}/2$ (SU(3)), and $\kappa=0$ (U(5)). For generic $\chi$, $H_{CQ}$ interpolates smoothly between these limits but does not itself correspond to an exact dynamical symmetry. The extended Hamiltonians used to describe triaxial rotational motion~\cite{vanisacker1986,zhang2022,zhang2024,teng2025,pan2024} go beyond this single-$\chi$ consistent-$Q$ framework by introducing additional, higher-order Casimir terms in $\hat{L}$ and $\hat{Q}_\chi$ (note, however, that the triaxial rotor mode of the BM collective model cannot produce $B_{4/2} < 1$ as mentioned above).
 
 Recent IBM studies have begun to address triaxial dynamics and their influence on anomalous  B(E2) behavior \cite{PhysRevC.111.014324}.
Calculations using a modified version of the IBM Hamiltonian that extended the SU(3)-symmetric modes by mapping a triaxial rotor onto the IBM Hamiltonian have been able to successfully reproduce $B(E2:4^+_1\rightarrow 2^+_1)/B(E2:2^+_1\rightarrow 0^+_{\mathrm{gs}}) < 1.0$  and $E(4^+_1)/E(2^+_1) > 2.0 $ for three cases: $^{168,170}$Os~\cite{zhang2022,zhang2024,teng2025,pan2024}  and $^{172}$Pt~\cite{zhang2022,pan2024}.  The static and dynamic aspects of the triaxial rotor were taken into account, including Hamiltonian terms of higher order in angular momentum, $\hat{L}$ and quadrupole moment, $\hat{Q}$~\cite{zhang2022} :  
 \begin{equation}
    H_{\text{Tri}} = \hat{H}_S + \hat{H}_D
\end{equation}
where
\begin{equation}
    \hat{H}_S = \frac{a_1}{N}  \hat{C}_2[\text{SU}(3)] + \frac{a_2}{N^3}  \hat{C}_2[\text{SU}(3)]^2 + \frac{a_3}{N^2 } \hat{C}_3[\text{SU}(3)],
\end{equation}
\begin{equation}
    \hat{H}_D = t_1 \hat{L}^2 + t_2 (\hat{L} \times \hat{Q} \times \hat{L})^{(0)} + t_3 (\hat{L} \times \hat{Q})^{(1)} \cdot (\hat{L} \times \hat{Q})^{(1)}.
\end{equation}
Here, $a_i$ and $t_i$ with $(i = 1, 2, 3)$ are real parameters that are determined by fitting to experimental spectra without explicit constraint on their signs. It was shown that the observed anomalous $B(E2)$ behavior arises when these terms interfere destructively. The SU(3) Casimir operators are defined as
\begin{equation}
    \hat{C}_2[\text{SU}(3)] = 2 \hat{Q} \cdot \hat{Q} + \frac{3}{4} \hat{L}^2,
\end{equation}
\begin{equation}
    \hat{C}_3[\text{SU}(3)] = -4 \sqrt{\frac{35}{9}} (\hat{Q} \times \hat{Q} \times \hat{Q})^{(0)}_0 - \sqrt{\frac{15}{2}} (\hat{L} \times \hat{Q} \times \hat{L})^{(0)}_0.
\end{equation}
In practice, the value of $\chi$ appearing in the expression for $\hat{Q}_\chi$ can be changed (within $\pm \sqrt{7} / 2$), going beyond the SU(3) limit. The three dynamical symmetry limits of the consistent-$Q$ Hamiltonian are characterized by: the $\mathrm{U}(5)$ limit for $\varepsilon>0$ and $\kappa=0$; the $\mathrm{O}(6)$ limit for $\varepsilon=0$, $\kappa<0$ and $\chi=0$; and the SU(3) limit for $\varepsilon=0, \kappa<0$ and $\chi= \pm \sqrt{7} / 2$. 
Triaxial-rotor calculations within the IBM-2 framework likewise reproduce suppressed $B_{4/2}$ ratios in selected W–Os–Pt isotopes~\cite{TENG2025139487}
While the extended IBM calculations that were shown to reproduce $B_{4/2}<1$ have been interpreted in terms of triaxial rotational motion, the corresponding Hamiltonians also incorporate additional degrees of freedom associated with SU(3) symmetry. The apparent success of triaxial-rotor descriptions may therefore not uniquely reflect triaxial deformation. Instead, it suggests that SU(3)-related correlations -- potentially unrelated to static triaxiality -- play a central role. This interpretation is further supported by the expectation that collective behavior near closed shells should initially manifest through vibrational, rather than rotational, degrees of freedom.

Anomalous $B_{4/2}<1$ behavior is also observed in lighter systems~\cite{kintish2014,tobin2014}. In particular, \textit{ab initio} no-core shell model calculations for $^{20}$Mg and $^{20}$O indicate significant contributions from SU(3) irreducible representations such as $(\lambda,\mu)=(4,2)$ and $(6,2)$ to low-lying yrast states~\cite{tobin2014}, consistent with similar observations in heavier nuclei including $^{48}$Cr and $^{114}$Te.
Complementary large-scale shell-model studies using random interactions have shown that collective-like spectra with $R_{4/2}\gtrsim 2$ and $B_{4/2}<1$ can emerge from strong mixing between ground-state and $\gamma$-band $4^+$ configurations~\cite{4qrd-1dqw}, pointing to a microscopic origin rooted in configuration mixing rather than conventional collective modes. Notably, nuclei exhibiting anomalous $B_{4/2}$ ratios in both the Te, Xe and W, Os, Pt regions share similar valence particle/hole configurations relative to the nearest closed shells ($N,Z=50$ and 82), suggesting an underlying valence-space symmetry that may be key to understanding the phenomenon.

\subsection{Mixed-symmetry interpretation}
The $B(E2)$ anomaly spans different regions of the nuclear chart. Here we focus on the W, Os, Pt ($A\sim170$) and Te, Xe ($A\sim110$) regions, which differ markedly in single-particle structure yet exhibit similar valence-boson numbers within the IBM framework. This suggests that the phenomenon is largely independent of specific orbital occupancies. Indeed, in the lighter systems neutrons occupy the $\nu h_{11/2}$ intruder orbital while protons populate the $\pi(d_{5/2},g_{7/2})$ subshells, whereas in the heavier region protons and neutrons occupy $\pi h_{11/2}$ and $\nu i_{13/2}$ orbitals, respectively. The persistence of anomalous $B_{4/2}$ values across these distinct configurations points to a more general mechanism.

Within the proton--neutron IBM-2, states are classified by the $F$-spin quantum number, with the fully symmetric ground-state band having $F=F_{\max}=(N_\pi+N_\nu)/2$ and the lowest mixed-symmetry states, involving out-of-phase proton--neutron motion (e.g. the well-known scissors mode~\cite{Iudice1978,bohle1984,Heyde2010}), having $F=F_{\max}-1$~\cite{iachello1984,vanisacker1986}. To explore whether coupling to such modes can account for the $B_{4/2}$ anomaly, we introduce an effective IBM-1 term, $H_{MS}$, constructed to mimic, upon mapping onto the IBM-2 space, the structural effect of admixing the lowest mixed-symmetry configuration into the yrast band. To explore this, we performed IBM calculations for Te and Xe isotopes using extended Hamiltonians of the form
\begin{equation}
    H=H_{CQ}+H_{MS},\end{equation}
where,
\begin{equation}
    H_{MS} = a\left(\hat{L}\times \hat{Q_{\chi}}\times \hat{L}\right)^0 +b\hat{L}\cdot\hat{L}.
\end{equation} 
We emphasize that $H_{MS}$, as written, acts within the single-boson-type IBM-1 space, in which $F$-spin is not itself a defined quantum number; the subscript ``MS'' denotes its effective role in reproducing the energetic and structural signature of mixed-symmetry admixture once interpreted at the IBM-2 level, rather than an exact $F$-spin label.

All Hamiltonian and transition-operator parameters used in the present calculations, together with the full IBM-1/IBM-2 codebase and example input files needed to reproduce Figs.~\ref{fig:E42_B42_exp_1} and \ref{fig:E42_B42_exp_2}, are available in the open-source pyIBM package~\cite{ahlgren_cederloef_pyibm_2025,cederlof2026pyibm}. Here the parameters are constrained by fits to  the experimental spectra and target transition ratios of the light Xe isotopes using a differential evolution algorithm with $L_2$ regularization. 
For $^{112}\text{Xe}$, the parameters (in keV, except for the dimensionless $\chi$) are $\chi = 0.50$, $\varepsilon = 350.31$, $\kappa = -21.09$, $a = -14.99$, and $b = 15.14$, yielding $B_{4/2} = 0.45 \pm 0.17$.
For $^{114}\text{Xe}$, the parameters are $\chi = 0.48$, $\varepsilon = 366.95$, $\kappa = -16.36$, $a = -13.00$, and $b = 14.34$, yielding $B_{4/2} = 0.70 \pm 0.08$.
For $^{116}\text{Xe}$ ($N=8$), the parameters are $\chi = 0.47$, $\varepsilon = 320.03$, $\kappa = -21.73$, $a = -8.20$, and $b = 12.58$, yielding $B_{4/2} = 1.55 \pm 0.05$. The calculations reproduce both spectra and suppressed $B_{4/2}$ values in $^{112,114}$Xe, although the latter remain sensitive to the relative strength of higher-order terms. The uncertainties in  $B_{4/2}$ values were estimated via numerical Jacobian covariance error propagation. 

Complementary large-scale shell-model (LSSM) calculations for $^{112}$Xe in the full $50$--$82$ valence space ($0g_{7/2}$, $1d_{5/2}$, $1d_{3/2}$, $2s_{1/2}$, $0h_{11/2}$) with the monopole-optimized effective interaction of Ref.~\cite{Qi2012} reproduce excitation energies but systematically yield $B_{4/2}\approx1.4$, close to the rotational limit. The calculations were carried out at the PDC Center for High Performance Computing, KTH. The inability of LSSM calculations to capture the anomaly, despite realistic configuration mixing, indicates that its origin is not purely single-particle but reflects a collective mechanism beyond standard configurations.

We interpret this behavior in terms of mixed-symmetry collective modes, as introduced above. These excitations exhibit strong proton and neutron quadrupole components with opposite phases, leading to partial cancellation in E2 matrix elements.

We propose that a low-lying mixed-symmetry quadrupole mode, admixed into the yrast structure, suppresses the $B(E2;4^+_1\rightarrow2^+_1)$ transition strength relative to $B(E2;2^+_1\rightarrow0^+_{\mathrm{gs}})$, thereby generating $B_{4/2}<1$. This mechanism naturally explains the coexistence of collective-like spectra ($R_{4/2}\gtrsim 2$) with reduced E2 strengths and is consistent with the enhanced role of neutron–-proton correlations in neutron-deficient systems, where valence protons and neutrons occupy similar orbitals.

The relevance of such modes is further supported by studies of light nuclei, where \textit{ab initio} calculations identify dominant SU(3) irreducible representations in systems exhibiting suppressed $B_{4/2}$~\cite{tobin2014}, as well as by shell-model studies showing that configuration mixing between ground-state and $\gamma$-band structures can produce similar anomalies~\cite{4qrd-1dqw}. The present results thus point to a unified interpretation of the $B(E2)$ anomaly in terms of mixed-symmetry collective dynamics rather than purely geometric deformation effects.

\section{Conclusions}
We conclude that the physical interpretation of the systematic appearance of ``anomalous'' cases of $B_{4/2} <1$ observed in two regions of the nuclear chart which has remained elusive until now, may be explained as due to the influence of mixed-symmetry collective excitations. This interpretation is supported by extended interacting boson model calculations including a more complete SU(3) symmetry. An extended IBM approach using third-order terms in angular momentum and quadrupole moment is to our knowledge currently the only theoretical model that successfully has reproduced the B(E2) anomaly. More detailed investigations are needed, both within the IBM framework and using alternative theoretical models including LSSM, PSM and beyond mean field models.

\section*{Acknowledgements}
This work was partly supported by the Swedish Research Council under Grant No. 2019-04808 (B.C.). CQ acknowledges financial support from the Olle Engkvist Foundation and the computational resource provided by PDC, KTH Royal Institute of Technology. The authors declare no competing interests.
\bibliography{main}
\bibliographystyle{epj}
\end{document}